\documentclass[letterpaper,11pt]{article}

\usepackage[utf8]{inputenc}
\usepackage{jheppub}

\title{Positivity of the Veneziano Amplitude in Ten Dimensions}
\author{Gareth Mansfield}
\affiliation{Mani L. Bhaumik Institute for Theoretical Physics\\Department of Physics and Astronomy\\
University of California, Los Angeles, CA 90095, USA}
\emailAdd{garethmansfield@physics.ucla.edu}

\usepackage[shortlabels]{enumitem}
\usepackage{amsmath}
\usepackage{amssymb}
\usepackage{amsthm}
\usepackage{graphicx}
\usepackage{physics}
\usepackage{fancyhdr}
\usepackage{pdflscape}

\makeatletter
\g@addto@macro\bfseries{\boldmath}

\makeatother

\def\N{\mathbb N}
\def\Z{\mathbb Z}

\def\C{\mathbb C}

\newcommand{\mc}{\mathcal}

\newtheorem{lemma}{Lemma}

\usepackage{hyperref}

\numberwithin{equation}{section}

\usepackage[svgnames]{xcolor}
\hypersetup{
    colorlinks,
    linkcolor={RoyalBlue},
    citecolor={RoyalBlue},
    urlcolor={RoyalBlue}
}

\abstract{The Veneziano amplitude describing the tree-level scattering of four open superstrings is expected to be consistent with unitarity in ten spacetime dimensions. While this follows indirectly from the no-ghost theorem, a direct proof at the level of the amplitude itself only exists for $D\leq 6$. In this article, we close this gap by providing a proof for the partial-wave positivity of the Veneziano amplitude in $D\leq 10$, derived directly from its representation as the Euler beta function. We also demonstrate that this proof can be modified to show the positivity of a wider family of amplitudes relevant to the $S$-matrix bootstrap. 
}

\begin{document}

\maketitle

\section{Introduction}

The Veneziano amplitude can be defined by the expression:
\begin{align}
    A^{\alpha_0}(s,t)=-(s+t)^{\alpha_0-1}\frac{\Gamma(-\alpha_0-s)\Gamma(-\alpha_0-t)}{\Gamma(-2\alpha_0-s-t)},\label{typeI}
\end{align}
where $\alpha_0\in \{0,1\}$ determines the Regge intercept. This famously satisfies a wide set of strict bootstrap constraints on tree-level four-particle amplitudes \cite{Veneziano:1968yb,Virasoro:1969me,Shapiro:1970gy}. It is crossing-symmetric in $s$ and $t$; it decays exponentially in the large-$s$, fixed-$t$ limit, implying tame UV behavior; it is meromorphic with only real, simple poles, implying consistency with causality; and it has polynomial residues,\footnote{In the $\alpha_0=0$ case, the residue at $s=0$ is not a polynomial in $t$. This can be amended by including other polynomial factors in the definition of $A^0$. In type-I string theory, this is achieved using the Yang-Mills numerator $\mc F(s,t)$.} implying consistency with locality. Due to these properties, the Veneziano amplitude has become a central object of study in the modern $S$-matrix bootstrap program \cite{Elvang:2015rqa,Cheung:2017pzi,Arkani-Hamed:2017jhn,Benincasa:2007xk,Cohen:2010mi}, and various ``deformations" of (\ref{typeI}) have been demonstrated to satisfy many of the same bootstrap constraints \cite{Coon:1969yw,Cheung:2022mkw,Geiser:2022exp, Geiser:2022icl,Geiser:2023qqq, Cheung:2023uwn,Bhardwaj:2023eus,Haring:2023zwu,Eckner:2024ggx,Huang:2022mdb,Cheung:2023adk}. Moreover, the amplitude $\mc A(s,t)=\mc F(s,t)A^{0}(s,t)$, where $\mc F(s,t)$ is the polynomial that appears in the Yang-Mills four-point numerator, also describes the tree-level scattering of four gluons in type-I string theory. Similarly, the amplitude $A^1(s,t)$ describes tree-level tachyon scattering in bosonic string theory. Due to this connection, much work has been done examining (\ref{typeI}) in efforts to directly bootstrap string theory \cite{Caron-Huot:2016icg,Berman:2023jys,Chiang:2023quf,Arkani-Hamed:2023jwn,Cheung:2024uhn,Cheung:2024obl,Berman:2024wyt,Albert:2024yap,Bhat:2024agd,Berman:2024kdh,Berman:2024eid,Berman:2024owc}.

Unitarity of the $S$-matrix imposes an additional constraint. When the exchanged state lies on-shell, the amplitude $A^{\alpha_0}(s,t)$ must factor into a product of 3-point amplitudes. Mathematically, this manifests as a requirement that the residue at each pole must admit a partial-wave decomposition,
\begin{align}
    \Res_{s=n}A^{\alpha_0}(s,t)=\sum_{j} B_{n,j} G^D_j\left(1+\frac{2t}{n+4\alpha_0}\right),\label{ansatz}
\end{align}
with the property that $B_{n,j}\geq 0$ for all  $n,j$. We refer to this constraint as \textit{positivity} of the amplitude. Here $G_j^D$ is the Gegenbauer polynomial for the spin-$j$ representation of $\text{O}(D-1)$.\footnote{In other contexts, this is notated as $C_j^{\alpha}$ with $\alpha=(D-3)/2$.}  Of particular relevance to the type-I ($\alpha_0=0$) Veneziano amplitude is the partial wave coefficient for exchange of a scalar $n=3$ state, which can be computed as
\begin{align}
    B_{3,0}^D=\frac{10-D}{24(D-1)}.
\end{align}
This expression becomes negative when $D> 10$, implying that consistency of the Veneziano amplitude with positivity forces $D\leq 10$.\footnote{It was shown in \cite{Arkani-Hamed:2022gsa} that positivity of the full $\mc A(s,t)=\mc F(s,t)A^0(s,t) $ amplitude in $D\leq 10$ is a direct consequence of the positivity of the residues at $n>0$ of the $A^0(s,t)$ part alone. Therefore, we may safely ignore the $\mc F(s,t)$ numerator for this analysis, and our results in this paper will still apply to both $A^0(s,t)$ and $\mc A(s,t)$.} Unexpectedly, this bound is in perfect agreement with the critical dimension of the type-I string, even though the amplitude itself possesses no $D$-dependence. So the tree-level amplitudes somehow ``know" about the critical dimension of the underlying string theory---a surprising fact that motivates one to ask whether positivity of all $B_{n,j}^{10}$ is, in fact, satisfied by the Veneziano amplitude.

Proving that every residue of (\ref{typeI}) can be rewritten in the form of (\ref{ansatz}) with $B_{n,j}^{D}\geq 0$ is challenging. Although this is technically possible and follows, in a rather indirect way, from the no-ghost theorem \cite{Goddard:1972iy}, the statement of positivity itself is relatively simple and demands a more direct and intuitive understanding---preferably one that does not require a detour through string theory. Such a proof could also provide new insight into the unitarity of various generalizations of the Veneziano amplitude, a problem which has attracted great interest in recent years \cite{Chakravarty:2022vrp,Bhardwaj:2022lbz,Figueroa:2022onw,Rigatos:2023asb,Jepsen:2023sia,Wang:2024wcc,Rigatos:2024beq, Bhardwaj:2024klc,Mansfield:2024wjc}. Some initial work on this problem was done in \cite{Maity:2021obe}, where the author proved the positivity of the $B_{n,n}^4$ coefficients of the bosonic ($\alpha_0=1$) Veneziano amplitude. More progress toward a direct proof was made in \cite{Arkani-Hamed:2022gsa}, which provided a complete proof of positivity of all $B_{n,j}^D$ for the type-I ($\alpha_0=0$) amplitude when $D\leq 6$, and the bosonic $(\alpha_0=1)$ amplitude when $D\leq 10$, by developing a new double-contour integral representation of the $B_{n,j}^D$ coefficients. More recently, further suggestive results have also been presented in \cite{Wang:2024wcc, Mansfield:2024wjc}. The Veneziano amplitude also seems to possess a stronger property than positivity, dubbed ``super-unitarity", which was recently conjectured and explored by \cite{Eckner:2024ggx}.  At the level of Wilson coefficients, the problem can also be reformulated as a set of bounds on certain polynomials of multiple zeta values, as shown in \cite{Green:2019tpt}. But so far, no direct proof of positivity in $D\leq 10$ has been presented in the literature.

In Section \ref{sec2}, we quickly review the previous work of \cite{Arkani-Hamed:2022gsa} that proved positivity of the type-I Veneziano amplitude in $D\leq 6$. We then extend the work of \cite{Arkani-Hamed:2022gsa} to a complete proof of positivity in $D\leq 10$. In Section \ref{sec3}, we demonstrate that this proof can be extended to some other amplitudes relevant to the $S$-matrix bootstrap: we explore the positivity of an interesting amplitude found in \cite{Cheung:2023adk}, and also utilize KLT relations to demonstrate positivity of some closed-string amplitudes.

\section{Positivity of the Veneziano Amplitude}\label{sec2}

\subsection[\texorpdfstring{Review of Positivity in $D\leq  6$}
                        {Review of Positivity in D ≤ 6}]{Review of Positivity in $D\leq 6$}

In this section, we reproduce the proof that the Veneziano amplitude is positive in $D\leq 6$ that was presented in \cite{Arkani-Hamed:2022gsa}, and show why this argument breaks down in $D>6$. This will only be a terse overview; a far more complete, detailed derivation and analysis is presented in the original paper. Following \cite{Arkani-Hamed:2022gsa}, we begin with the expression of $A^0(s,t)$ as the analytic continuation of the integral,
\begin{align}
    A^0(s,t)=\frac{\Gamma(-s)\Gamma(-t)}{\Gamma(1-s-t)}=\frac1s\int_0^1 \dd z\> z^{-s} (1-z)^{-1-t},\label{worldsheet}
\end{align}
and use the Cauchy integral formula to expand out the Taylor coefficients of the integrand: 
\begin{align}
    (1-z)^{-1-t}=\sum_{k=0}^\infty \frac{1}{k!}\left[\oint_{y=0}\frac{\dd y}{2\pi i}\frac{(1-y)^{-1-t}}{y^{k+1}}\right] z^k.\label{taylor}
\end{align}
Substituting (\ref{taylor}) into (\ref{worldsheet}), then exchanging the sum and integral, allows for straightforward computation of the residue. When $n\geq 1$, we find that
\begin{align}
    \Res_{s=n}A^0(s,t)=\frac1{n!}\oint_{y=0}\frac{\dd y}{2\pi i}\frac{(1-y)^{-1-t}}{y^{n}}.
\end{align}
For the sake of simplicity, we will suppress all manifestly positive prefactors like $1/n!$ from here on. Rewrite $t$ in terms of the scattering angle $x=\cos\theta=1+\frac{2t}{n}$, and substitute $y$ with a variable $u$ such that $1-y=e^{-u}$, to obtain
\begin{align}
    R_n(t)=\oint_{u=0}\frac{\dd u}{2\pi i}\frac{e^{\frac{un}{2}(x+1)}}{(e^u-1)^{n}}.\label{eq:residue_contour}
\end{align}
Now we would like to compute the $j$th Gegenbauer coefficient of the residue at $s=n$, which is given by the orthogonality relation
\begin{align}
    B_{n,j}^D=\int_{-1}^1\dd x\> (1-x^2)^{\frac{D-4}{2}}G_j^{(D)}(x)R_n(t)=\int_{-1}^1\dd x\> (1-x^2)^{J}\pdv[j]{x}R_n(t)
\end{align}
where we have simplified the expression using the Rodrigues formula for $G_j^{(D)}$, and also defined $J= (D-4)/2+j$. The $j$th derivative of (\ref{eq:residue_contour}) can be evaluated explicitly, which allows us to compute the partial wave coefficients as 
\begin{align}
    B_{n,j}=\oint_{u=0}\frac{\dd u}{2\pi i}\frac{u^j e^{un/2}}{(e^u-1)^{n}}\int_{-1}^1\dd x\>(1-x^2)^{J}e^{xun/2}.\label{hard}
\end{align}
The integral over $x$ in (\ref{hard}) does not have a neat closed form, but following the procedure of \cite{Arkani-Hamed:2022gsa}, there is a useful way it can be rewritten: if we restrict to $D\in 2\Z$, then we can expand the integrand via the following identity:
\begin{align}\begin{split}
    (1-x^2)^{J}e^{\beta x}&= (- x\pm 1)^{J} e^{\mp \beta
     }\pdv[J]{\beta}e^{\beta ( x\pm 1)}.
     \end{split}
\end{align}
Commuting this derivative through the integral and manipulating the bounds of integration, we get the following form:
\begin{align}
   \int_{-1}^1\dd x\>(1-x^2)^{J}e^{\beta x}=(-1)^J\left[e^{\beta
    } \pdv[J]{\beta}e^{-2\beta}\int_0^\infty - e^{-\beta
    } \pdv[J]{\beta} e^{2\beta}\int_0^\infty \right]  y^{J}e^{\beta y}\dd y,
\end{align}
and the remaining integral here is the Laplace transform of $y^J$. Computing it and plugging this back into (\ref{hard}) obtains
\begin{align}\begin{split}
    B_{n,j}= &\oint_{u=0}\frac{\dd u}{2\pi i}\frac{u^j}{(e^u-1)^{n}} \pdv[J]{u} \frac{e^{un}}{u^{J+1}}
    - \frac{u^je^{un}}{(e^u-1)^{n}} \pdv[J]{u} \frac{e^{-un}}{u^{J+1}}.
\end{split}\end{align}
When $n+j$ is even, the first and second term differ by an overall sign, so the integrand vanishes and we are left with $B_{n,j}=0$. If $n+j$ is odd, it can be shown that the first and second terms here are equivalent. Then we can integrate by parts $J$ times, rewrite the derivative with respect to $u$ as a second contour integral, and simplify to obtain
\begin{align}
B_{n,j}^D=\oint_{u=0}\frac{\dd u}{2\pi i}\oint_{v=0}\frac{\dd v}{2\pi i}\frac{(v-u)^j}{(uv)^{\frac{D-2}{2}+j}(e^{v}-e^{u})^{n}}.
\end{align}
From here, we make a change of variables $1-x=e^u$, $1-y=e^v$ to rewrite this as
\begin{align}
     B_{n,j}=c_{n,j}^D\oint_{x=0}\frac{\dd x}{2\pi i}\oint_{y=0}\frac{\dd y}{2\pi i}&\frac{(1-x)^{-1}(1-y)^{-1}}{(\log(1-x)\log(1-y))^{\frac{D-2}{2}}} \frac{\left(\frac{1}{\log(1-x)}-\frac1{\log(1-y)}\right)^{j}}{(x-y)^{n}}\label{contour}.
\end{align}
Here we have also restored the positive prefactor $c_{n,j}^D$. The formula (\ref{contour}) is the central result of \cite{Arkani-Hamed:2022gsa}. Examining this formula, positivity in $D=6$ follows from the fact that the Laurent expansions of the functions $(1-z)^{-1}/\log(1-z)^2$ and $(\log(1-x)^{-1}-\log(1-y)^{-1})^{j}/{(x-y)^{n}}$ both only contain positive coefficients; therefore the coefficient on $x^{-1}y^{-1}$ of their product must be positive. This in turn implies positivity in all $D\leq 6$: from the branching rules for representations of $\text{SO}(D)$, it follows that any $D$-dimensional Gegenbauer polynomial can be decomposed into a positive linear combination of $(D-1)$-dimensional Gegenbauer polynomials. Therefore positivity in $D$ dimensions implies positivity in $D-1$ dimensions.

Lastly, we can see that this method fails if $D>6$: taking $D=8$, we find that 
\begin{align}
    \frac{(z-1)^{-1}}{\log(1-z)^3}=z^{-3}-\frac12 z^{-2}+\frac z{240}+\text{other positive terms.}
\end{align}
The introduction of just a single negative term ($-z^{-2}/2$) in this series spoils the proof. The issue becomes more severe in $D=10$, as the expansion of $(1-z)^{-1}/\log(1-z)^4$ contains eight negative terms. However, in the next section we will show that we can transform the integrand into one where all coefficients are positive by adding a second term with vanishing residue.

\subsection[\texorpdfstring{Positivity in $D\leq 10$}%
                        {Positivity in D ≤ 10}]{Positivity in $D\leq 10$}\label{sec:d10}

In this section, we make a modification to the contour formula (\ref{contour}) and use this to prove the positivity of the Veneziano amplitude in ten dimensions. To begin, we introduce a notation that mimics that of \cite{Arkani-Hamed:2022gsa}, by writing $B_{n,j}^D=c_{n,j}^D\beta_{n,j}^D$, where $\beta_{n,j}^D$ is the double-contour integral on the right side of (\ref{contour}). Then in $D=10$, this evaluates to
\begin{align}\begin{split}
    \beta_{n,j}^{10}=\oint_{x=0}\frac{\dd x}{2\pi i}\oint_{y=0}\frac{\dd y}{2\pi i}&\frac{(1-x)^{-1}(1-y)^{-1}}{(\log(1-x)\log(1-y))^{4}}\frac{\left(\frac{1}{\log(1-x)}-\frac1{\log(1-y)}\right)^j}{(x-y)^n}.\label{contour10}
\end{split}\end{align}
By direct computation up to order $\mc O(z^7)$, we see that the Laurent expansion of $(1-z)^{-1}\log(1-z)^{-4}$ contains at least eight negative coefficients. We need a way to cancel out the contribution of these terms to the residue. To accomplish this, for each $j$ we introduce a rational function
\begin{align}
    P_{j}(x,y)=\frac{1}{(xy)^j}\left[\sum_{k=-4}^{6}\frac{x^{k}}{y^{4}}+\frac{y^{k}}{x^{4}}+\sum_{k=-2}^{1}\frac{jx^{k}}{y^{3}}+\frac{jy^{k}}{x^{3}}\right],\label{rational}
\end{align}
and observe that the residue of $P_{n,j}(x,y)/(x-y)^{n-j}$ at $y=x=0$ vanishes if $n\geq 8$. Therefore we may add this term to the integrand of (\ref{contour10}) without changing the integral's value, obtaining
\begin{align}\begin{split}
    \text{For }n\geq 8:\>\>\>\beta_{n,j}=&\oint_{x=0}\frac{\dd x}{2\pi i}\oint_{y=0}\frac{\dd y}{2\pi i}\>\frac{1}{(x-y)^{n-j}}\\&\times\Bigg[P_{j}(x,y)+\frac{(1-x)^{-1}(1-y)^{-1}}{(\log(1-x)\log(1-y))^{4}}\left(\frac{\frac{1}{\log(1-x)}-\frac1{\log(1-y)}}{x-y}\right)^j\Bigg].\label{extraterm}
\end{split}\end{align}
From here it suffices to argue that for sufficiently large $j$, the factor in brackets admits a Laurent expansion with exclusively nonnegative coefficients. Since the Laurent coefficients of $(x-y)^{j-n}$ are also nonnegative, this would imply that the integrand is a product of two nonnegative Laurent series, which would then guarantee the non-negativity of the resulting residue. First, to make the notation more concise, we define Laurent coefficients $Q_{\mu\nu}^j$ as
\begin{align}
  \frac{(1-x)^{-1}(1-y)^{-1}}{(\log(1-x)\log(1-y))^{4}}\left(\frac{\frac{1}{\log(1-x)}-\frac1{\log(1-y)}}{x-y}\right)^j=\frac{1}{(xy)^j}\sum^\infty_{\mu,\nu=-4}Q^j_{\mu,\nu} x^\mu y^\nu.\label{sum}
\end{align}
Now we can explicitly compute the values of $Q^j_{\mu\nu}$ at $-4\leq \mu\leq 6$, $-4\leq \nu\leq 6$ in order to verify that any negative coefficients in this set are canceled out by the function $P_j(x,y)$ as we desired. Explicitly, we find for $j\geq 1$ that
\begin{align}\begin{split}
    \sum^\infty_{\mu,\nu=-4}Q^j_{\mu,\nu} x^\mu y^\nu=&- \frac1{x^3} - \frac1{720} - \frac{x}{720} - \frac{x^2}{945} - \frac{11 x^3}{15120} 
 - \frac{47 x^4}{103680} - \frac{19 x^5}{80640} \\
 &- \frac{439 x^6}{6842880} -\frac{j+4}{24x^2}y-\frac{j}{720x}y-\frac{j-2}{1440}y-\frac12\frac{j-7}{5040}xy\\
 &\Bigg.+(x\leftrightarrow y)+(\text{positive terms})+\mc O(x^7)+\mc O(y^7).
 \end{split}
\end{align}
Each negative term listed here has magnitude less than or equal to the corresponding term of equal degree in $P_j(x,y)$, and therefore cannot cause (\ref{extraterm}) to dip below zero. Therefore it suffices to prove that all the remaining $Q^j_{\mu\nu}$ coefficients at $\mu>6$ or $\nu>6$ are nonnegative. Specifically, the remainder of this section is devoted to proving the following claim:
\begin{center}
    \textit{When $j\geq 4$, if $\mu>6$ or $\nu> 6$, then $Q^j_{\mu,\nu}\geq 0$.}
\end{center}
This statement is sufficient to prove that $\beta_{n,j}\geq 0$ when $n\geq 8$ and $j\geq 4$. The positivity of the remaining $\beta_{n,j}$ coefficients at $n\leq 7$ can be verified by explicit computation, and the positivity of the coefficients with $j\leq 3$ is verified by the method outlined in Section 4.2 of \cite{Arkani-Hamed:2022gsa}. To prove the claim, we will induct on $j$. To streamline the proof, the base case of $j=4$ has been relegated to Section \ref{sec:basecase}. Proceeding with the inductive step, we assume that $Q^{j}_{\mu\nu}\geq 0$ for all coefficients with $\mu\geq 7$ or $\nu\geq 7$, and seek to prove this statement for $Q^{j+1}_{\mu\nu}$. We define another set of coefficients $G_k$ from the following Laurent expansion:
\begin{align}
    \frac{z}{\log(1-z)}=-1+\sum_{k=1}^{\infty}G_kz^k.
\end{align}
Here, $G_k$ is equivalent to the absolute value of the $k$th Gregory coefficient \cite{Jordan:1947}. In terms of these coefficients, it will be useful to expand
\begin{align}\begin{split}
    \frac{\frac1{\log(1-x)}-\frac1{\log(1-y)}}{x-y}&=-\frac{x^{-1}-y^{-1}}{x-y}+\sum_{m=0}^\infty G_{m+1} \frac{x^m-y^m}{x-y}\\
    &=\frac1{xy}+\sum_{m=1}^\infty\sum_{k=0}^{m-1} G_{m+1} x^{m-1-k}y^k.\label{expand}
\end{split}\end{align}
Substituting (\ref{expand}) into (\ref{sum}) allows us to express $Q^{j+1}_{\mu\nu}$ as a convolution of the $Q^{j}$ and $G_k$ coefficients, obtaining the following recursive formula:
\begin{align}
Q^{j+1}_{\mu,\nu}=Q^{j}_{\mu,\nu}+\sum_{\sigma=-4}^{\mu-1} \sum_{\rho=-4}^{\nu-1}Q^{j}_{\sigma,\rho}G_{\mu+\nu-\sigma-\rho}.\label{recur}
\end{align} 
Many of the terms in this sum are positive by the inductive hypothesis. Discarding these terms provides a lower bound
\begin{align}
    Q^{j+1}_{\mu,\nu}\geq \sum_{\sigma=-4}^{6}\sum_{\rho=-4}^{\min(\nu-1,6)}Q^{j}_{\sigma,\rho} G_{\mu+\nu-\sigma-\rho},\label{recur2}
\end{align}
where we have assumed $\mu\geq 7$ without loss of generality. To check the positivity of this sum, we will need the following lemma, which gives a stronger characterization of the $G_k$ coefficients.

\begin{lemma}

\noindent \textit{The $G_k$ coefficients satisfy the following bounds:\label{lemma1}
\begin{enumerate}[(a),ref={1(\alph*)}]
    \item For any polynomial $p(x)=\sum_k r_kx^{k}$, if $p(x)$ is positive almost everywhere on $x\in[0,1]$, then $\sum_k r_k G_k> 0$.\label{lemma1a}
    \item $\sum_{k\leq k_0} r_kG_{k+\alpha}>0$ for all $\alpha\in\N$ if 
    \begin{align}
        \sum_{k\leq k_0} r_k
    (\theta(r_k) G_{k_0}+\theta(-r_k)G_k)>0,
    \end{align}
    where $\theta$ denotes the unit step function.\label{lemma1b}
\end{enumerate}
}
\end{lemma}

\noindent \textit{Proof.} To begin, we can interpret $G_k$ as the following residue:
\begin{align}
    G_k=\Res_{z=0}\frac{1}{z^{k}\log(1-z)}=\Res_{x=\infty}\frac{x^{k-2}}{\log(1-\frac1x)}.
\end{align}
This function has a simple pole at infinity and a branch cut on $x\in [0,1]$, and it is holomorphic everywhere else on $\C$. We can evaluate the residue by integrating clockwise along a contour that encloses the branch cut:
\begin{align}
    G_k=\int_0^1 \dd x'  \>\text{Disc}_{x=x'}\>\frac{x^{k-2}}{\log(1-\frac1x)}.
\end{align}
Evaluating the difference between the two branches gives the following formula for $G_k$:\footnote{Under a change of variables $x\to\frac{1}{1+x}$, this reproduces Schröder's integral formula for $G_k$ \cite{Schroder}.}
\begin{align}
    G_{k}=\int_0^1 \frac{x^{k-2}\dd x}{\log(\frac1x-1)^2+\pi^2}.\label{Hausdorff}
\end{align}
The formula (\ref{Hausdorff}) gives that $G_k$ is a Hausdorff moment sequence, meaning that we have found a positive measure $\dd\mu$ on $[0,1]$ for which $G_k=\int_0^1 x^k\dd\mu$. The remaining components of the proof are general bounds which hold for any sequence with this property. To see (a), we observe that any finite linear combination of $G_k$ can be written as
\begin{align}
    \sum_k r_kG_k=\int_0^1 \left(\sum_{k} r_k x^{k}\right)\frac{x^{-2}\dd x}{\log(\frac1x-1)^2+\pi^2}.
\end{align}
If $p(x)=\sum_{k} r_k x^k$ is positive almost everywhere on the interval $x\in[0,1]$, then this integral must be positive, which proves (a). To prove (b), we first note that choosing $p(x)=x^k(1-x)$ implies that $G_{k}/G_{k+1}>1$. Next, for each $N\in\N$ we define the $N\times N$ Hankel matrix $G^N$ with components $G^N_{i,j}=G_{i+j}$. Then we observe that for any vector $v\in\C^N$, we can compute the matrix product
\begin{align}
    v^*G^Nv=\sum_{i,j}G_{i+j}\overline v_i v_j=\int_0^1 \left|\sum_{k} v_k x^{k}\right|^2\frac{x^{-2}\dd x}{\log(\frac1x-1)^2+\pi^2},
\end{align}
which is positive if $v\neq 0$, so $G^N$ is a positive definite linear operator on $\C^N$. Therefore, all its principal square submatrices have positive determinants, which then implies $G_{k+1}/G_k>G_{k}/G_{k-1}$. Using these new results, we construct the bound
\begin{align}\begin{split}
    \sum_{k\leq k_0}r_k G_{k+\alpha}&= G_{k_0+\alpha}\sum_{k\leq k_0}r_k \left(\theta(r_k)\frac{G_{k+\alpha}}{G_{k_0+\alpha}}+\theta(-r_k)\frac{G_{k+\alpha}}{G_{k_0+\alpha}}\right)\\&\geq  G_{k_0+\alpha}\sum_{k\leq k_0}r_k \left(\theta(r_k)+\theta(-r_k)\frac{G_{k}}{G_{k_0}}\right),\end{split}
\end{align}
which is equivalent to the claim. \qed

\vspace{6pt}

Now we can verify that the sum (\ref{recur2}) is positive by using lemma \ref{lemma1a}: for any $\nu\neq -3$, under a substitution $G_k\to x^k$ in (\ref{recur2}) we obtain a polynomial in $j$ and $x$ that, upon inspection, is positive on the region $x\in(0,1]$, $j\geq 4$. In the case $\nu=-3$, we make the same argument, but need to take the upper limit of the sum to be $\sigma= 12$ in order for the relevant polynomial in $x$ to be positive; this is unproblematic as we can directly check that the remaining coefficients $Q^j_{7,-3},\ldots,Q^j_{12,-3}$ are all given by manifestly positive polynomials in $j$. This concludes the proof of the inductive step; all that remains is to show that the claim holds in the base case $j=4$.

\subsection[\texorpdfstring{Positivity of $Q^4_{\mu\nu}$}{Positivity of Q at j = 4}]{Positivity of $Q^4_{\mu\nu}$}\label{sec:basecase}

\vspace{6pt}

In this section, we prove the base case of the induction in Section \ref{sec:d10} which states that $Q_{\mu\nu}^4\geq 0$ when $\mu\geq 7$ or $\nu\geq 7$. Unfortunately, the proof of this statement is more computationally heavy than the previous sections. We first need to obtain a set of bounds on the $Q_{\mu\nu}^j$ coefficients at arbitrary $\mu$ with particular fixed values of $\nu$. Note that the coefficients are symmetric ($Q_{\mu\nu}^j=Q_{\nu\mu}^j$), so these bounds also apply to $Q_{\mu\nu}^j$ at arbitrary $\nu$ with fixed $\mu$. 

\noindent \begin{lemma}\label{lemma2}
    
\textit{The coefficients defined in (\ref{sum}) obey the following bounds:
\begin{enumerate}[(a),ref={2(\alph*)}]
    \item $Q^j_{k,-4}\geq 0$ when $k\geq 7$.\label{lemma2a}
    \item $Q^4_{k,-3}\geq 0$ when $k\geq 7$.\label{lemma2b}
    \item $G_4Q^j_{k,-2}+G_3Q^j_{k,-3}+G_4Q^j_{k,-4}\geq 0$ when $1\leq j\leq 4$.\label{lemma2c}
    \item $Q^j_{k,i}\geq 0$ when $k\geq 7$, $-2\leq i\leq 6$, and $1\leq j\leq 4$.\label{lemma2d}
\end{enumerate}}
\end{lemma}

\noindent \textit{Proof.} We follow a strategy similar to one used in \cite{Arkani-Hamed:2022gsa} to prove the positivity of various Laurent series. We begin by writing out a contour-integral representation of the $Q^j_{\mu,\nu}$ coefficients:
\begin{align}
    Q_{\mu\nu}^j=\oint_{x=0}\frac{\dd x}{2\pi i}\oint_{y=0}\frac{\dd y}{2\pi i}\frac{1}{x^{\mu+1}y^{\nu+1}}\frac{(1-x)^{-1}(1-y)^{-1}}{\log(1-x)^4\log(1-y)^4}\left(\frac{\frac{1}{\log(1-x)}-\frac1{\log(1-y)}}{x-y}\right)^j.
\end{align}
For fixed values of $\nu$, the integral over $y$ can be evaluated explicitly. Then for $\nu\in\{-4,-3,-2\}$ we find the following generating functions for the $Q_{\mu\nu}^j$ coefficients:
\begin{align}\begin{split}
    Q^j_{\mu,-4}& =\oint_{x=0}\frac{\dd x}{2\pi i}\frac{(1-x)^{-1}x^{-1-\mu}}{\log(1-x)^4},\\
    Q^j_{\mu,-3} &=\oint_{x=0}\frac{\dd x}{2\pi i}\frac{ (2x-2j+jx) \log(1 - x)-2 j x}{2 (-1 + x) \log(1 - x)^5 x^{2+\mu}},\\
    Q^j_{\mu,-2}&=\oint_{x=0}\frac{\dd x}{2\pi i}\frac{ 12 (j-1) j x^2 -
     12 j x (j (x-2) + x) \log(
       1 - x)}{24 (1-x) x^{3 + \mu} \log(1 - x)^6} \\&\qq{}\qq{}+ \frac{12 j (1 + j) - 
        12 j (2 + j) x + (1 + j) (4 + 3 j) x^2}{24 (1-x) x^{3 + \mu} \log(1 - x)^4}.\end{split}\label{intformulas}
\end{align}
As the value of $\nu$ increases, the resulting integrals become increasingly complex. Beginning with statement (a), we will rewrite the formula for $Q_{\mu,-4}^j$ in (\ref{intformulas}) using the integral expansion
\begin{align}\begin{split}
    \frac{(1-z)^{-1}z^4}{\log(1-z)^4}=\frac16\int_0^1&\dd s\>(1 - z)^{s - 1} \Big[(s-1)^3 z^3+(3s^2-9s+7) z^2 + (6s-12) z + 6\Big].\end{split}\label{intformula}
\end{align}
We may now evaluate the contour integral by differentiating each side of (\ref{intformula}) $k$ times at $z=0$, which yields the expression
\begin{align}
\begin{split}
    Q^j_{k-4,-4}=\frac1{6(k!)}\int_0^1\dd s\> (1 - s)_{(k-3)}\Big[&k^3 s^3 - k^2 s(6s^2+3s+1)\\
    &+ks(11s^2+15s+7)-s(6s^2+18s+12)\Big],\label{eqone}
   \end{split}
\end{align}
where $a_{(n)}=a(a+1)\ldots (a+n-1)$ denotes the rising factorial. Treating the factor in brackets as a polynomial in $s$ and $k$ over the domain $s\in[0,1]$, $k\geq 7$, we can bound it from below by a much simpler polynomial:
\begin{align}
    Q^j_{k-4,-4}> \frac1{6(k!)}\int_0^1\dd s\> (1 - s)_{(k-3)}\Big[&k^3 s^3 - 11 k^2\Big].
\end{align}
The integrand is negative when $s\in[0,\sqrt[3]{11/k}]$ and positive elsewhere. Using this fact, we may split the integral into positive and negative components, and bound their sum from below by
\begin{align}\begin{split}
   Q^j_{k-4,-4}> \frac{(k-3)!}{6k!}\int_0^{\sqrt[3]{11/k}} &\dd s\> (k^3 s^3 - 11 k^2)\\&+\frac{(k-2)!}{6k!}\int_{\sqrt[3]{11/k}}^1 \dd s\> (1-s)\left(k^3 s^3 - 11 k^2\right).
\end{split}
\end{align}
This evaluates to a rational function of $\sqrt[3]{k}$ which, upon inspection, is positive when $k\geq 31$. Positivity in the remaining cases $7\leq k\leq 30$ can then be verified by direct computation. Next, to prove (b) we employ the same general procedure. We follow the same procedure as in part (a) to find a representation of the form
\begin{align}\begin{split}
    Q^j_{k-4,-3}= \frac{1}{24 k!} \int_0^1
\dd s \>(1-&s)_{(k-3)}\Big[ (3 - k) (16 + jk-4k) s + 2 (2 - k) (18 + jk-6k) s^2\\ &+ 
 2 (2 - k) (1 - k) (6 + jk-2k) s^3 - j (2 - k) (1 - k) k s^4\Big].\label{eqtwo}
 \end{split}
\end{align}
When $j= 4$, this admits the following simpler lower bound:
\begin{align}\begin{split}
     Q^j_{k-4,-3}\geq \int_0^1
\dd s \>(1-s)_{(k-3)}(4s^{2}k-88).\end{split}
\end{align}
From here we use the same argument as in part (a): we prove that this integral is positive for sufficiently large $k$, and then manually verify positivity at small $k$. This completes the proof of (b). Next, bound (c) can be proven by taking the appropriate linear combinations of the expressions in (\ref{intformulas}) and then applying the same procedure that was done for (a) and (b), while taking extra care to ensure the proof holds at arbitrary $j$. Since this proof follows the same procedure we have already written, and requires significantly longer calculations, we will omit it here. Lastly, bound (d) can also be proven by repeatedly applying this method nine more times; again, we will not explicitly write this here. However, in Appendix \ref{appA} we instead present a ``shortcut" to proving (d) that significantly reduces the number of required calculations. \qed

\vspace{6pt}

Equipped with these bounds, the most efficient way to demonstrate positivity of the remaining coefficients at $\mu,\nu\geq 7$ is to induct on $j$ as we did in Section \ref{sec:d10}, but this time to begin at $j=0$. Then we will terminate the induction once it reaches $j=4$.

\begin{lemma}
    \label{lemma3}\textit{If $\mu\geq 7$, $\nu\geq 7$, and $j\leq 4$, then $Q^j_{\mu\nu}\geq 0$.}
\end{lemma} 

\noindent \textit{Proof.} The case of $j=0$ follows immediately from observing the factorization $Q^0_{\mu\nu}=Q^0_{\mu,-4}Q^0_{-4,\nu}$ and then applying lemma \ref{lemma2a} to demonstrate the positivity of each factor when $\mu,\nu\geq 7$. Proceeding by induction on $j$, assume that $Q^{j}_{\mu\nu}\geq 0$ for all coefficients with $\mu\geq 7$ and $\nu\geq 7$. Now we recall the recursive formula of (\ref{recur}):
\begin{align}
Q^{j+1}_{\mu,\nu}=Q^{j}_{\mu,\nu}+\sum_{\sigma=-4}^{\mu-1} \sum_{\rho=-4}^{\nu-1}Q^{j}_{\sigma,\rho}G_{\mu+\nu-\sigma-\rho}.
\end{align} 
The sum in this formula can be decomposed into four pieces:
\begin{align}
    Q^{j+1}_{\mu,\nu}=Q^{j}_{\mu,\nu}+\left[\sum_{\sigma,\rho=-4}^{6} + \sum_{\sigma=7,\rho=-4}^{\mu-1,6} + \sum_{\sigma=-4,\rho=7}^{6,\nu-1} +\sum_{\sigma,\rho=7}^{\mu-1,\nu-1}\right]Q^{j}_{\sigma,\rho} G_{\mu+\nu-\sigma-\rho}.
\end{align}
As we argued in Section \ref{sec:d10}, both $Q^{j}_{\mu,\nu}$ and the final sum in brackets are nonnegative by the inductive hypothesis, and the leftmost sum in brackets is positive by lemma \ref{lemma1a}. To show positivity of the two remaining terms, we break into two cases. If $j\geq 1$, then we may use that the inductive hypothesis, together with lemmas \ref{lemma2a} and \ref{lemma2d}, implies that all terms in each sum, other than $Q_{\sigma,-3}^j$ and $Q_{-3,\rho}^j$, are positive. So we may safely assume that there is no negative contribution from any other terms. Then to show positivity of the sum of these remaining terms, by lemma \ref{lemma1b} it is sufficient to have nonnegativity of the sum
\begin{align}
G_{4}Q^{j}_{\sigma,-2}+G_{3}Q^{j}_{\sigma,-3}+G_{4}Q^{j}_{\sigma,-4},
\end{align}
and this is the content of lemma \ref{lemma2c}. In the final remaining case of $j=0$, we instead must make use of the factorization $Q^{0}_{\sigma,\rho}=Q^0_{\sigma,-4}Q^0_{-4,\rho}$ to rewrite 
\begin{align}
    \sum_{\sigma=7}^{\mu-1}Q_{\sigma,\rho}^0G_{\mu+\nu-\sigma-\rho}=\sum_{\rho=-4}^{6}Q^0_{\sigma,-4}\left(\sum_{\sigma=7,\rho=-4}^{\mu-1,6}Q_{-4,\rho}^0G_{\mu+\nu-\sigma-\rho}\right)
\end{align}
We know that $Q^0_{\sigma,-4}\geq 0$ for $\sigma\geq 7$ by lemma \ref{lemma2a}. Positivity of the sum in parenthesis can be seen from the inequality
\begin{align}
   Q^{0}_{-4,-4} G_{12}+Q^{0}_{-4,-3} G_{11}+Q^{0}_{-4,-2} G_{12}+\sum_{k=0}^6 G_{8-k}Q^{0}_{-4,k}>0,\label{mid1}
\end{align}
which can be verified by direct calculation. Then we can apply lemma \ref{lemma1b} in order to generalize (\ref{mid1}) to arbitrary $G_k$, which gives the desired bound. This completes the proof. \qed

\vspace{6pt}

From here, the positivity of $Q_{\mu\nu}^4$ is immediate. When $j= 4$, lemmas \ref{lemma2a}, \ref{lemma2b}, and \ref{lemma2d} guarantee the positivity of $Q^4_{\mu\nu}$ when $\nu< 7$ and $\mu\geq 7$, and when $\nu\geq 7$ and $\mu< 7$. Then lemma \ref{lemma3} guarantees that $Q^4_{\mu\nu}\geq 0$ when both $\mu\geq 7$ and $\nu\geq 7$. Combining these results, we obtain our final desired bound of $Q_{\mu\nu}^4\geq 0$ when $\mu\geq 7$ or $\nu\geq 7$. This proves the base case of the induction in Section \ref{sec:d10}, and therefore completes the proof of the positivity of $\beta_{n,j}$.

\section{Other Amplitudes}\label{sec3}

\subsection{Virasoro-Shapiro}
Here we will apply the result of Section \ref{sec:d10} to prove the positivity of the Virasoro-Shapiro amplitudes for type-II and heterotic strings in ten dimensions. To start, we recall that the Veneziano amplitudes for open type-I ($\alpha_0=0$) and bosonic ($\alpha_0=1$) strings are given respectively by
\begin{align}
    A^{0}(s,t)=\frac{\Gamma(-s)\Gamma(-t)}{\Gamma(1-s-t)}&&A^{1}(s,t)=-\frac{\Gamma(-1-s)\Gamma(-1-t)}{\Gamma(-2-s-t)},
\end{align}
and their residues are
\begin{align}
    \text{Res}_{s=n}A^{0}(s,t)=\frac{(t+1)_{(n-1)}}{n!}&&\text{Res}_{s=n}A^{1}(s,t)=\frac{(t+2)_{(n+1)}}{(n+1)!}
\end{align}
where $a_{(n)}=a(a+1)\ldots (a+n-1)$. In the previous section, we proved the positivity of $A^{0}(s,t)$ in ten dimensions, and the positivity of $A^{1}(s,t)$ in $D=10$ has already been proven in \cite{Arkani-Hamed:2022gsa}. Our goal now is to extend these results to closed-string amplitudes, which can be constructed from the open ones using the KLT relations \cite{Kawai:1985xq}. This has already been done for the bosonic string in \cite{Arkani-Hamed:2022gsa}: the double copy relation gives the bosonic Virasoro-Shapiro amplitude as
\begin{align}
    A^{\text{Bosonic}}(s,t)= \frac{\sin(\pi s)\sin(\pi t)}{\pi\sin(\pi(s+t))}A^{1}(s,t)^2=\frac{\Gamma(-1-s)\Gamma(-1-t)\Gamma(-1-u)}{\Gamma(2+s)\Gamma(2+t)\Gamma(2+u)}
\end{align}
and its residues are given by 
\begin{align}
    \Res_{s=n} A^{\rm Bosonic}(s,t)&=\left(\frac{(t+2)_{(n+1)}}{(n+1)!}\right)^2
\end{align}
This is simply the square of the residue for the open bosonic string, which we know to be positive in $D=10$ by \cite{Arkani-Hamed:2022gsa}. The product of two positive linear combinations of Gegenbauer polynomials gives another positive linear combination of Gegenbauer polynomials; this can be seen by expanding the $j\otimes j'$ representation of $\text{SO}(D-1)$ down into a positive sum over irreps. Then as explained in \cite{Arkani-Hamed:2022gsa}, this guarantees positivity of the bosonic Virasoro-Shapiro amplitude in ten dimensions. But now we can employ our new result from Section \ref{sec:d10} to extend this argument to closed superstrings. The type-II Virasoro-Shapiro amplitude is defined by
\begin{align}
     A^{\rm II}(s,t)= \frac{\sin(\pi s)\sin(\pi t)}{\pi\sin(\pi(s+t))}A^{0}(s,t)^2=\frac{\Gamma(-s)\Gamma(-t)\Gamma(-u)}{\Gamma(1+s)\Gamma(1+t)\Gamma(1+u)},
\end{align}
while the Virasoro-Shapiro amplitude for heterotic strings is given by
\begin{align}
     A^{\rm Heterotic}(s,t)=\frac{\sin(\pi s)\sin(\pi t)}{\pi\sin(\pi(s+t))}A^{0}(s,t)A^{1}(s,t)=\frac{\Gamma(-s-1)\Gamma(-t-1)\Gamma(-u+3)}{\Gamma(1+s)\Gamma(1+t)\Gamma(1+u)}.
\end{align}
From each of these, we can compute the residues at each pole $s\in \N$:
\begin{align}\begin{split}
    \Res_{s=n}A^{\rm II}(s,t)&=\left(\frac{(t+1)_{(n-1)}}{n!}\right)^2\\
    \Res_{s=n} A^{\rm Heterotic}(s,t)&=\frac{(t+1)_{(n-1)}(t+2)_{(n+1)}}{n!(n+1)!}
\end{split}
\end{align}
For the type-II string, the residue is simply the square of the residue for the type-I string, which we have shown to be positive; therefore the type-II Virasoro-Shapiro amplitude is positive in ten dimensions. For the heterotic string, the residue can be expanded as the product
\begin{align}\begin{split}
    \Res_{s=n} A^{\rm Heterotic}(s,t)&=\frac{\left(t+n-1\right)_{(4)}}{(n-2)_{(4)}}\Res_{s=n}A^{0}(s,t)\Res_{s=n-4}A^{1}(s,t).
    \end{split}
\end{align}
We note that it is necessary to take the residue of $A^1(s,t)$ at $s=n-4$ rather than $n$, to correct for the difference in the scattering angle $\cos\theta=1+2t/(n+4\alpha_0)$ between the $\alpha_0=0$ and $\alpha_0=1$ amplitudes. Now we have obtained a product involving two residues that we know to admit positive Gegenbauer expansions in $D=10$, and a degree-four polynomial that can also be shown to admit a positive Gegenbauer expansion. This guarantees the positivity of the heterotic string amplitude in ten dimensions.

\subsection[\texorpdfstring{The Massless Hypergeometric Amplitude in $D=5$}{The Massless Hypergeometric Amplitude in  D = 5}]{The Massless Hypergeometric Amplitude in $D=5$}

Another amplitude of interest to the $S$-matrix bootstrap is defined as
\begin{align}
    A^{\text{HG}}(s,t)=-\frac1s {}_3F_2\left[\begin{matrix}1,-s,\frac12+t\\1-s,\frac12\end{matrix};1\right]\label{d5}
\end{align}
where ${}_3F_2$ is the generalized hypergeometric function \cite{alma991033988579703276}. This arises as a special case of the more general hypergeometric amplitude studied in \cite{Cheung:2023adk}, and is notable in part because, in five spacetime dimensions, unitarity prevents us from deforming its mass gap away from zero. That is, if we consider an arbitrary shift of the external momenta $s\to s+\alpha_0$, $t\to t+\alpha_0$ by some real number $\alpha_0$, then a mixture of analytical and numerical results \cite{Cheung:2023adk,Mansfield:2024wjc} expects that positivity holds if and only if $\alpha_0=0$. In this section, we will prove the reverse direction of this claim by showing that if $\alpha_0=0$, then $B_{n,j}^5\geq 0$ for all $n,j\in\N$. Additionally, this amplitude is noteworthy because its partial wave coefficients arise (up to manifestly positive factors) in the $\lambda\to 0$ limit of the ``planar analogue amplitude" studied in \cite{Cheung:2024obl}:
\begin{align}
A^{\rm planar}(s, t)&=-\frac{\rho}{s t}  +\frac{\lambda \Gamma(\lambda-\lambda s) \Gamma(\lambda-\lambda t)}{\Gamma(2 \lambda-\lambda s-\lambda t)}{ }_3 F_2\left[\begin{array}{c}
\lambda(1-s), \lambda(1-t), \frac{3 \lambda-1}{2} \\
\lambda(2-s-t), \frac{3 \lambda+1}{2}
\end{array} ; 1\right],\label{planar}
\end{align}
Thus, here we also guarantee the $D=5$ critical dimension in the $\lambda\to 0$ limit of (\ref{planar}) that was conjectured by \cite{Cheung:2024obl}. We intend this as a demonstration of the broader applicability of our method; we leave for future work a complete, systematic generalization of this proof to other amplitudes that admit contour formulas similar to (\ref{contour}).

It was shown in \cite{Mansfield:2024wjc} that the partial wave coefficients of both (\ref{d5}), and the $\lambda\to 0$ limit of (\ref{planar}), admit a contour integral representation that looks similar to the analogous formula for Veneziano:
\begin{align}
    \beta_{n,j}^{D,\text{HG}}=\oint_{x=0}\frac{\dd x}{2\pi i}\oint_{y=0}\frac{\dd y}{2\pi i}&\frac{(1-x)^{-1/2}(1-y)^{-1/2}}{(\log(1-x)\log(1-y))^{\frac{D-2}{2}}} \frac{\left(\frac{1}{\log(1-x)}-\frac1{\log(1-y)}\right)^{j}}{(x-y)^{n}}.\label{contourHG}
\end{align}
Because the contour formula is only valid at even $D$, we cannot simply substitute $D=5$. It will instead suffice to show that all $B_{n,j}$ coefficients other than those at $n=2$ are positive in $D\leq 6$, as we can quickly verify by direct calculation that $B^5_{2,0}=B^5_{2,1}=0$ and $B^5_{2,2}>0$. With this in mind, we proceed in the same way as we did for Veneziano, and introduce a new term
\begin{align}
    P_j(x,y)=\frac1{(xy)^j}\sum_{k=-1}^{12} \frac{x^k}{y^2}+\frac{y^k}{x^2}.
\end{align}
We observe that $P_j(x,y)/(x-y)^{n-j}$ has vanishing residue at $y=x=0$ when $n>12$, so for $n>12$ we may add it to the integrand of (\ref{contourHG}):
\begin{align}\begin{split}
    \beta^{5,\text{HG}}_{n,j}=\oint_{x=0}\frac{\dd x}{2\pi i}\oint_{y=0}\frac{\dd y}{2\pi i}&\>\frac{1}{(x-y)^{n-j}}\\\times\Bigg[P_{j}(x,y)
    &+\frac{(1-x)^{-1/2}(1-y)^{-1/2}}{(\log(1-x)\log(1-y))^{2}}\left(\frac{\frac{1}{\log(1-x)}-\frac1{\log(1-y)}}{x-y}\right)^j\Bigg].
    \end{split}
\end{align}
The positivity of the remaining coefficients at $n\leq 12,n\neq 2$ can be verified by hand. Define the $Q^j_{\mu\nu}$ coefficients as
\begin{align}
  \frac{(1-x)^{-1/2}(1-y)^{-1/2}}{(\log(1-x)\log(1-y))^{2}}\left(\frac{\frac{1}{\log(1-x)}-\frac1{\log(1-y)}}{x-y}\right)^j=\frac{1}{(xy)^j}\sum^\infty_{\mu,\nu=-2}Q^j_{\mu,\nu} x^\mu y^\nu,
\end{align}
and following the a similar argument to what we did for Veneziano, it suffices to show that for all $j\geq 1$, $Q_{\mu\nu}^j\geq 0$ when $\mu\geq 13$ or $\nu\geq 13$. Beginning with the base case of $j=1$, we can expand out 
\begin{align}
Q^1_{\mu\nu}=Q^0_{\mu\nu}+\sum_{\sigma,\rho=-2}^{\mu-1,\nu-1} Q^0_{\sigma\rho}G_{\mu+\nu-\sigma-\rho}.
\end{align} 
as was done in (\ref{recur}). If $\mu, \nu\geq 13$, then we can decompose the sum as
\begin{align}
Q^1_{\mu\nu}=Q^0_{\mu\nu}+\left(\sum_{\sigma,\rho=-2}^{12}+\sum_{\sigma=13,\rho=-2}^{\mu-1,12}+\sum_{\sigma=-2,\rho=13}^{12,\nu-1}+\sum_{\sigma,\rho=13}^{\mu-1,\nu-1}\right) Q^0_{\sigma\rho}G_{\mu+\nu-\sigma-\rho}.
\end{align}
Now we note the factorization $Q_{\mu\nu}^0=Q_{\mu,-2}^0Q_{-2,\nu}^0$. Then via the same process used in lemma \ref{lemma2a}, we can prove that $Q_{\mu,-2}^0\geq 0$ when $\mu\geq 13$, so then $Q_{\mu\nu}^0\geq 0$ when $\mu,\nu\geq 13$. This gives positivity of both $Q^{0}_{\mu\nu}$ and last term in parentheses. Next, positivity of the first term in parentheses can be shown by demonstrating positivity of the polynomial
\begin{align}
f(x)=\sum_{\sigma,\rho=-2}^{12}    Q^0_{\sigma\rho}x^{\mu+\nu-\sigma-\rho}
\end{align}
for $x\in(0,1]$ and applying lemma \ref{lemma1a}. Positivity of the remaining second and third terms follows from the factorization
\begin{align}
    \left(\sum_{\sigma=13,\rho=-2}^{\mu-1,12}+\sum_{\sigma=-2,\rho=13}^{12,\nu-1}\right)Q^0_{\sigma\rho}G_{\mu+\nu-\sigma-\rho}=2\sum_{\sigma=13}^{\mu-1}Q^0_{\sigma,-2}\left(\sum_{\rho=-2}^{12}Q^0_{-2,\rho}G_{\mu+\nu-\sigma-\rho}\right)\label{factorization}
\end{align}
and deferring to lemma \ref{lemma1b}: to prove positivity of the sum in parentheses in  (\ref{factorization}), it suffices for the expression
\begin{align}
   G_{15}+\sum_{\rho=-1}^{12}Q^0_{-1,\rho}G_{13-\rho},
\end{align}
to be positive. This can be verified by direct computation. Now it still remains to verify positivity in the case $\mu>12,\,\nu\leq 12$ and $\nu>12,\,\mu\leq 12$. This could be done rather laboriously using a contour integral representation similar to (\ref{intformulas}) for each of the fifteen $Q^j_{\mu\nu}$ coefficients that need to be computed, or it can be done by a shortcut that mimics the one discussed in Appendix \ref{appA}. In either case, this procedure is a repetition of the one used for the Veneziano amplitude, so we will not write it out explicitly here. Now we proceed by induction and assume that $Q^{j}_{\mu,\nu}\geq 0$ when $\mu\geq 13$ or $\nu\geq 13$. We can decompose $Q^{j+1}_{\mu\nu}$ into the following sum:
\begin{align}
    Q^{j+1}_{\mu\nu}=Q^{j}_{\mu\nu}+Q^{j}_{-2,-2} G_{4+\mu+\nu}+2\sum_{\sigma=-1}^{\mu-1}Q_{\sigma,-2}^jG_{2+\mu+\nu-\sigma}+\sum_{\sigma,\rho=-1}^{\mu-1}Q^{j}_{\sigma\rho}G_{\mu+\nu-\sigma-\rho}.
\end{align}
All terms are positive by the inductive hypothesis or by direct evaluation, save for the sum involving $Q^j_{\sigma,-2}$. For this sum, we can defer to lemma \ref{lemma1b} and argue that positivity of this term follows from the positivity of the sum
\begin{align}
    G_{28}+2\sum_{\sigma=-1}^{12}Q_{\sigma,-2}^jG_{2+12+12-\sigma}.\label{last}
\end{align}
Since $Q^j_{\sigma,-2}=Q^0_{\sigma,-2}$, the value of this sum is independent of $j$; therefore (\ref{last}) can be verified directly by hand. This completes the proof.

\section{Discussion}

We have proven that the tree-level Veneziano amplitude for the scattering of four type-I strings satisfies partial-wave positivity in $D\leq 10$, without deferring to the no-ghost theorem. This was accomplished by extending a contour-integral representation of the partial wave coefficients that was developed by \cite{Arkani-Hamed:2022gsa} and had originally led to a proof of positivity in $D\leq 6$. We expect that the method could be generalized to demonstrate positivity of any amplitude that admits a double-contour integral representation similar to (\ref{contour}); such representations are known to exist for the hypergeometric \cite{Mansfield:2024wjc}, Coon \cite{Bhardwaj:2022lbz}, and bosonic Veneziano \cite{Arkani-Hamed:2022gsa} amplitudes. 
We have demonstrated one instance of this use case by modifying our proof to confirm a conjecture that the hypergeometric amplitude constructed in \cite{Cheung:2023adk}, when evaluated at $(r,m^2)=(-\frac12,0)$, satisfies positivity in $D\leq 5$. We have also extended our result to closed strings by using the KLT relations to demonstrate positivity of the Virasoro-Shapiro amplitude for type-II and heterotic string theory in $D\leq 10$.

The proofs we have presented for both the Veneziano and hypergeometric amplitude share an interesting property in that they operate by defining some finite $j_0$, proving the positivity of $B_{n,j}$ analytically for all $j\geq j_0$, and then resorting to a manual proof of positivity for each $0\leq j<j_0$ by explicit computation. This reflects a conjecture that has appeared numerous times in the literature for various generalized string amplitudes \cite{Bhardwaj:2022lbz, Wang:2024wcc,Rigatos:2024beq,Mansfield:2024wjc}, which loosely states that the positivity condition is controlled exclusively by the $B_{n,j}^D$ coefficients at low values of $j$. Our proof reflects this fact in the $D=10$ Veneziano and $D=5$ hypergeometric cases and could therefore act as a starting point toward developing a more rigorous statement and proof of this conjecture. This would allow for the development of a more complete, systematic analysis of unitarity constraints on other amplitudes relevant to the $S$-matrix bootstrap.

It would be worthwhile to try to develop a single, simpler formula for $B_{n,j}^D$ from which positivity in $D=10$ is manifest, in the same way that (\ref{contour}) makes the positivity of $B_{n,j}^6$ manifest. This could further demystify the surprising positivity condition of the Veneziano residues and could also alleviate practical difficulties in extending our analysis to other amplitudes. In particular, it is difficult to demonstrate the positivity of the $\alpha_0=1$ Veneziano amplitude for the bosonic string in its critical dimension $D=26$. In attempting to prove this, we find that the base case of the induction is far more difficult to prove than it was for the type-I amplitude, and requires significantly more computation. Therefore, a more practical method to carry out this proof would be desirable. Lastly, as remarked in \cite{Arkani-Hamed:2022gsa}, it would also be interesting to explore possible deeper origins of the contour formula (\ref{contour}) itself, and in particular to find a better way of understanding its curious $u\leftrightarrow v$ symmetry.

\section*{Acknowledgements}
The author thanks Nima Arkani-Hamed, Zvi Bern, Rishabh Bhardwaj, Shounak De, Lorenz Eberhardt, Sebastian Mizera, Grant Remmen, and Marcus Spradlin for many helpful discussions. This material is based upon work supported by the National Science Foundation Graduate Research Fellowship under Grant No. DGE-2444110. The author is also grateful to the Mani L. Bhaumik Institute for Theoretical Physics for support.

\appendix

\section[\texorpdfstring{Proof of the Low-Order $Q^j_{\mu\nu}$ Bound}{Proof of the Low-Order Q Bound}]{Proof of the Low-Order $Q^j_{\mu\nu}$ Bound}

\label{appA}
In this section, we return to lemma \ref{lemma2d}, which states

\vspace{6pt}

\noindent \textbf{Lemma 2(d).} \textit{$Q^j_{\mu,\nu}\geq 0$ when $\mu\geq 7$, $1\leq j\leq 4$, and $-2\leq \nu\leq 6$.}

\vspace{6pt}

As mentioned in Section \ref{sec:basecase}, one could prove this statement on a case-by-case basis for each $-2\leq \nu\leq 6$ by repeating the procedure of lemmas \ref{lemma2a} and \ref{lemma2b} a total of nine times. In this section, we will show that it is sufficient to do this procedure only at $\nu=-2$, and that all remaining cases follow from these. The proof follows the same general structure as of lemma \ref{lemma3}, with some modifications to account for the different restriction on $\nu$.

\vspace{6pt}

\textit{Proof.} Positivity of $Q^j_{\mu,-2}$ when $\mu\geq 7$ needs to be proven using the integral representation in (\ref{intformulas}); the following proof is not valid for this coefficient. For the remaining coefficients, beginning with the base case $j=1$, positivity of $Q_{\mu\nu}^1$ when $\mu=7$ and $-1\leq \nu\leq 6$ can be verified by hand. Then when $\mu\geq 8$ and $-1\leq \nu\leq 6$, we expand as
\begin{align}\begin{split}
    Q^{1}_{\mu\nu}=\sum_{\sigma,\rho=-4}^{6,\nu-1}Q_{\sigma\rho}^{0} G_{\mu+\nu-\sigma-\rho}&+ \sum_{\sigma=7,\rho=-4}^{\mu-2,\nu-1}Q_{\sigma\rho}^{0} G_{\mu+\nu-\sigma-\rho}\\&+\left[Q^{0}_{\mu,\nu}+\sum_{\rho=-4}^{\nu-1}Q^{0}_{\mu-1,\rho}G_{1+\nu-\rho}\right].
    \end{split}
\end{align}
For the first term, one can quickly check that the polynomial
\begin{align}
    f(x,j)=\sum_{\sigma,\rho=-4}^{6,\nu-1} Q_{\sigma\rho}^{0} x^{\mu+\nu-\sigma-\rho}\label{square}
\end{align}
is positive on the region $x\in(0,1)$, $j\geq 0$, for each $-1\leq \nu\leq 6$, which gives positivity by lemma \ref{lemma1a}. Positivity of the second term follows from noting the factorization $Q^0_{\sigma\rho}=Q^0_{\sigma,-4}Q^0_{-4,\rho}$, and rewriting as
\begin{align}
    \sum_{\sigma=7,\rho=-4}^{\mu-2,\nu-1}Q_{\sigma\rho}^{0} G_{\mu+\nu-\sigma-\rho}=\sum_{\sigma=7}^{\mu-2}Q_{\sigma,-4}^{0}\left( \sum_{\rho=-4}^{\nu-1}Q_{-4,\rho}^{0} G_{\mu+\nu-\sigma-\rho}\right).\label{Qeq}
\end{align}
We can explicitly check that $Q_{-4,-4}^0=1$, $Q_{-4,-2}^0=\frac16$, and all remaining $Q^0_{-4,\rho}$ in (\ref{Qeq}) are negative. Therefore we may use the positivity of the sum
\begin{align}
 Q^{0}_{-4,-4} G_{12}+Q^{0}_{-4,-3} G_{11}+Q^{0}_{-4,-2} G_{12}+\sum_{k=0}^{\nu-1} G_{8-k}Q^{0}_{-4,k}>0
\end{align}
for all $-1\leq \nu\leq 6$ to prove positivity via lemma \ref{lemma1b}. Finally, to show positivity of the third term, we may factor it as
\begin{align}
    Q^0_{\mu-1,-4}\left[\frac{Q^0_{\mu,-4}}{Q^0_{\mu-1,-4}}Q^0_{-4,\nu}+\sum_{\rho=-4}^{\nu-1}Q^0_{-4,\rho} G_{2+\nu-\rho}\right].\label{qwerty}
\end{align}
Then using that $0<\frac{Q^0_{\mu,-4}}{Q^0_{\mu-1,-4}}< 2$, which can be proven by a modification to lemma \ref{lemma2a}, the positivity of this sum can be seen by directly evaluating (\ref{qwerty}) at each $-1\leq \nu\leq 6$. Now we proceed by induction on $j$ until we reach $j=4$. Assume that $Q^j_{\mu\nu}\geq 0$ when $\mu\geq 7$ and $-2\leq \nu\leq 6$. Then we may expand out $Q^{j+1}_{\mu\nu}$ as:
\begin{align}
    Q^{j+1}_{\mu\nu}=\sum_{\sigma,\rho=-4}^{\mu-1,\nu-1}Q_{\sigma\rho}^{j} G_{\mu+\nu-\sigma-\rho} 
\geq \sum_{\sigma,\rho=-4}^{6,\nu-1} Q_{\sigma\rho}^j 
\end{align}
where the inequality is obtained by dropping all terms that are positive due to the inductive hypothesis and lemma \ref{lemma2c}. Positivity of remaining sum follows the positivity of the polynomial (\ref{square}) together with lemma \ref{lemma1a}. So we have shown that the entire sum is positive, completing the proof. \qed

\bibliographystyle{JHEP}
\bibliography{main}

\end{document}